\documentclass[10pt]{iopart}
\usepackage{graphicx}
\usepackage{amssymb}
\usepackage[margin=2cm]{geometry}

\begin{document}

\title{Magnetic field generation from a coil-shaped foil by a laser-triggered hot-electron current}

\author{A. V. Brantov, Ph. Korneev, V. Yu. Bychenkov}
\address{P. N. Lebedev Physics Institute, Russian Academy of
Science, Leninskii Prospect 53, Moscow 119991, Russia}

\vspace{10pt}
\begin{indented}
\item[]December 2018
\end{indented}

\begin{abstract}
A strong electron current triggered by a femtosecond relativistically
intense laser pulse in a foil coil-like target is shown to be able to
generate a solenoidal-type extremely strong magnetic field. The magnetic
field lifetime sufficiently exceeds the laser pulse duration and is defined
mainly by the target properties. The process of the magnetic field
generation was studied with 3D PIC simulations. It is demonstrated that the
pulse and the target parameters allow controlling the field strength and
duration. The scheme studied is of great importance for laser-based
magnetization technologies.
\end{abstract}



\section{Introduction}\label{sec1}

Optical approaches are known to provide efficient and robust sources of
extremely high magnetic fields in a laser laboratory. One of the main
mechanisms responsible for quasistatic magnetic field production in such
schemes is the discharge currents in a specially shaped conducting material,
for example, metal wires \cite{Korobkin-stpl79, Santos-pop2018, korneev-njp17}.
But with very intense relativistic laser pulses, an additional effect comes
from the current of accelerated electrons \cite{Korneev-pre15}. The
laser-accelerated electrons leaving a target form a direct current, which
generates a strong azimuthal magnetic field around it. If the laser pulse is
very short, then the discharge process occurs as a propagating discharge
wave \cite{Quinn-prl2009}; a shaping of the discharge conductor can result
in a sophisticated time-dependent field structure \cite{Kar-nc2016}. Such
optically generated magnetic fields are very interesting for radiation and
particle acceleration, particle guiding, and a wide range of other
magnetized plasma studies.

Here, we present a scheme that under certain conditions utilizes both the
current of the laser-accelerated electrons and the discharge current. In our
case with a single-turn foil loop, they form a closed current contour
similarly to what the discharge current does in capacitor-coil schemes
\cite{Korobkin-stpl79, Santos-pop2018}. The scheme can be easily adjusted to
control the magnetic field properties by changing the target geometry and
the laser parameters. Using a very compact geometry compared to
capacitor-coil targets \cite{Korobkin-stpl79, Santos-pop2018}, for example,
allows using very short laser pulses to confine the generated propagating
discharge electromagnetic wave. Along with large petawatt laser facilities,
this allows applying this scheme with table-top femtosecond lasers.

\section{PIC simulation of magnetic field generation.}

\begin{figure}[!ht]
\includegraphics[width=5.5cm]{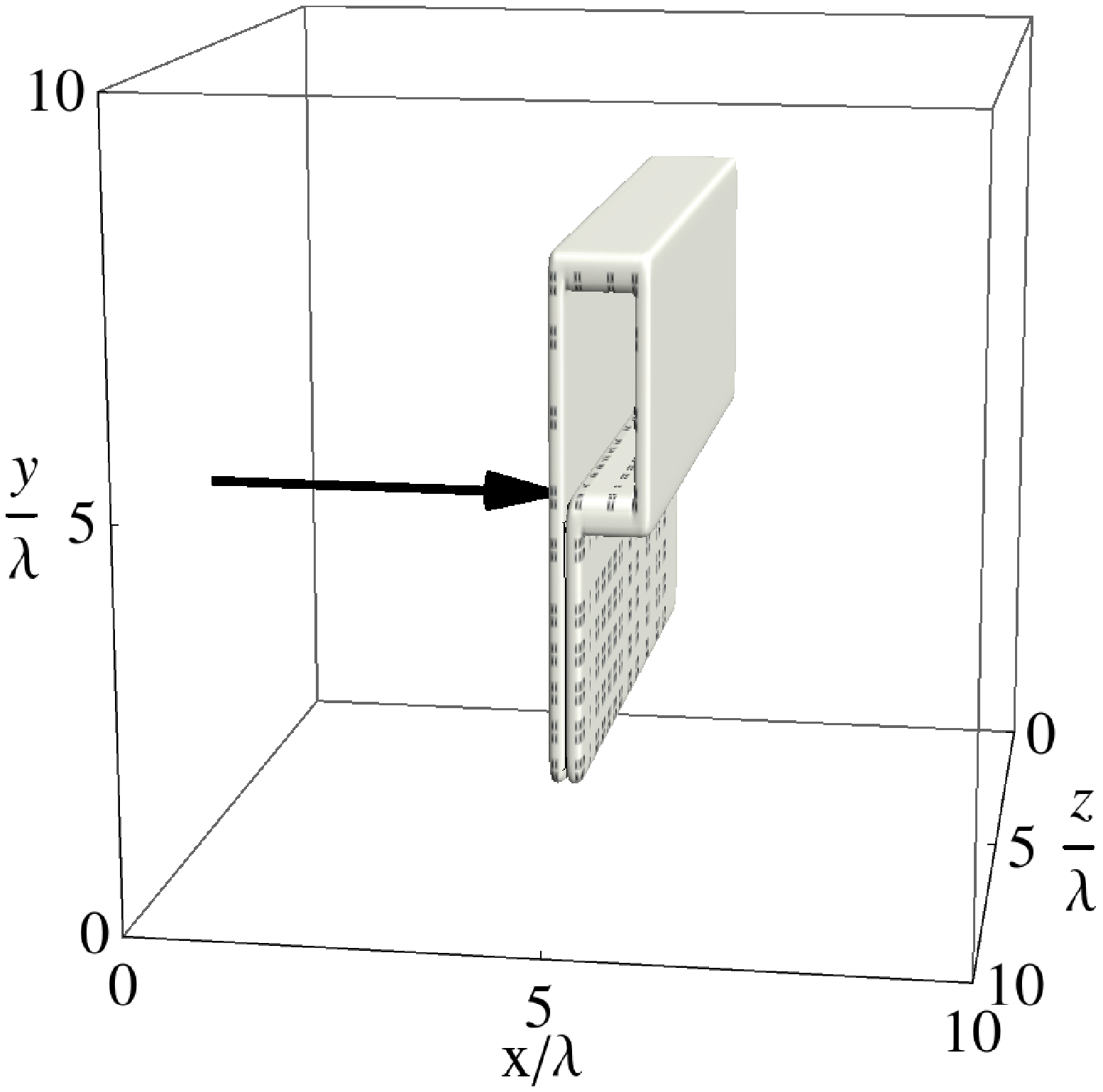}\hspace{1.5cm}
\includegraphics[width=5.5cm]{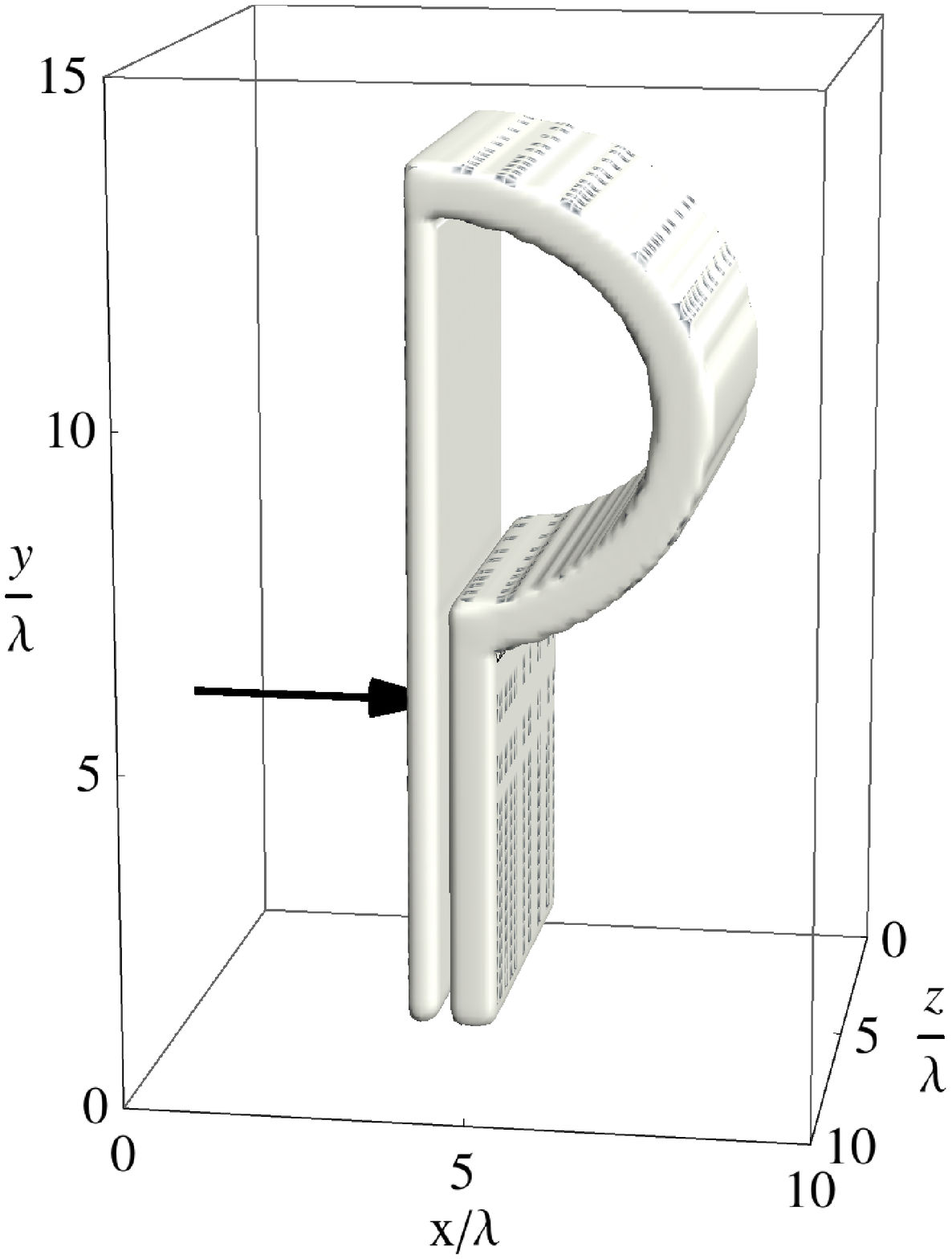}
\caption{Targets used in PIC simulations: the black arrows indicate the
laser beam axis.}
\label{fig1}
\end{figure}

The 3D simulation of magnetic field generation by a shot laser pulse was
performed with the PIC code Mandor. A laser pulse of 0.5\,J energy has a
Gaussian shape in time (30\,fs FWHM duration) and space (4\,$\mu$m FWHM
focal spot size). These parameters correspond to a maximum pulse intensity
$10^{20}$\,W/cm$^2$, which gives a dimensionless linearly polarized laser
field amplitude $a_0=8.55$ (for the laser wavelength $\lambda=1\,\mu$m).
We used two type of targets (see Fig.~\ref{fig1}) consisting of electrons
with density of $50 n_c$, where $n_c$ is critical density, and heavy
immovable ions. The simulations were collisionless and were performed with
spatial grid steps of $0.01\lambda\times0.02\lambda\times0.05\lambda$ in a
simulation box $X\times Y\times Z=10\lambda\times15\lambda\times10\lambda$.
In all the cases, the laser pulse was focused at the front side of the
target surface in a hot spot with the center point coordinates
$(X{=}5,\,Y{=}5,\,Z{=}5)$.

The less time-consuming simulation with the first target design was used
simply to demonstrate the basic principle of magnetic field generation, and
the second allows increasing the volume of the strong magnetic field,
although both target designs are basically the same. This design uses the
foil emitter--collector setup, where the laser-irradiated flat face part of
the foil (emitter) emits hot electrons, which bombard the flat plate
(collector) behind. There is a narrow vacuum gap 1\,$\mu$m thick between the
plates. The plates are actually parts of a shaped foil with the same
thickness making a single-turn loop (see Fig.~\ref{fig1}).

The idea is to localize the current of fast electrons by a certain initially
defined target shape and to add it to a discharge current excited by the
laser charging of the irradiated plate \cite{Korobkin-stpl79}. This
superposition increases the efficiency of the scheme over a scheme with only
the discharge current. The compact size of the target is consistent with the
femtosecond laser pulse duration, although it can also switch the regime of
magnetic field generation. In the considered situation, the magnetic field
would be a field of an electromagnetic discharge wave \cite{Kar-nc2016}, not
a field of a long quasistatic discharge current \cite{Santos-pop2018}.

\begin{figure}[!ht]
\includegraphics[width=6.5cm]{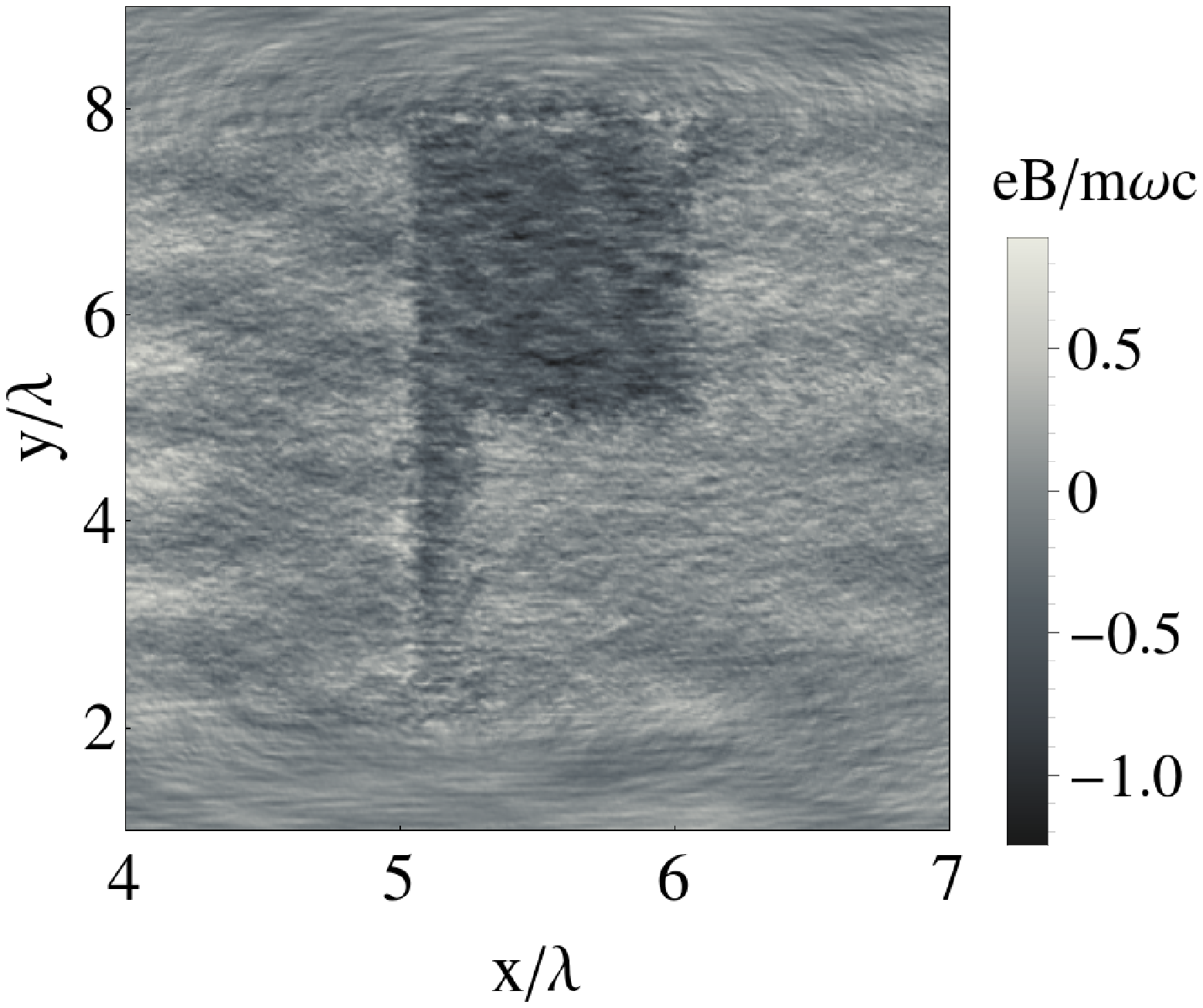}\hspace{1.5cm}
\includegraphics[width=6.5cm]{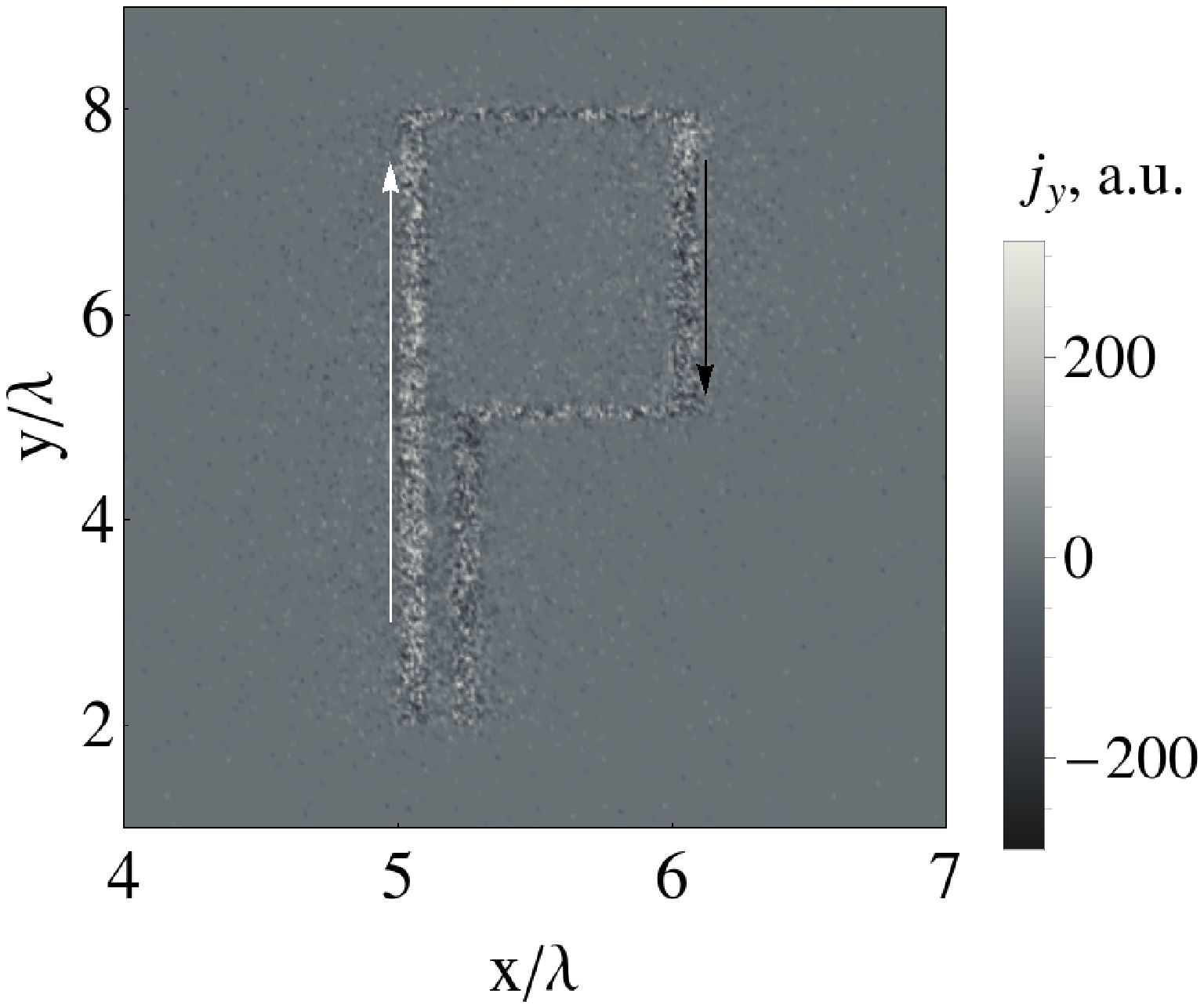}
\caption{The magnetic field $B_z$ (left panel) and electron current $j_y$
(right panel) just after the laser pulse action ($t=87$\,fs): the arrows
(black and white) in the right panel indicate the current direction.}
\label{fig2}
\end{figure}

\begin{figure} [!ht]
\includegraphics[width=7 cm]{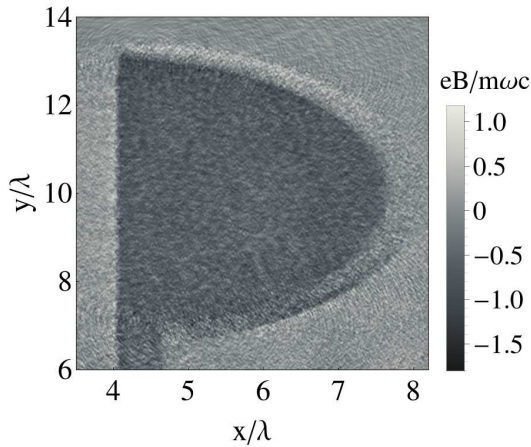}
\caption{The magnetic field $B_z$ just after the laser pulse action
($t=87$\,fs): the dark black color corresponds to negative $B_z$ values.}
\label{fig3}
\end{figure}

The results of PIC simulations for the first target are shown in Fig.~\ref{fig2}.
It is clearly seen that a negative magnetic field is generated in the free
internal volume of the loop (see the left panel in Fig.~\ref{fig2}). This
field is due to positive components of electron current on the left target
side and its negative component on the right target side, which is also seen
in the right panel in Fig.~\ref{fig2}.
The electron current corresponds to electrons moving toward the interaction
volume. This current is formed by the evacuation of hot electrons under the
action of the short laser pulse. To increase magnetic field volume, we
passed to the second target type, which also demonstrates formation of a
strong magnetic field (see Fig.~\ref{fig3}).

\begin{figure} [!ht]
\includegraphics[width=8 cm]{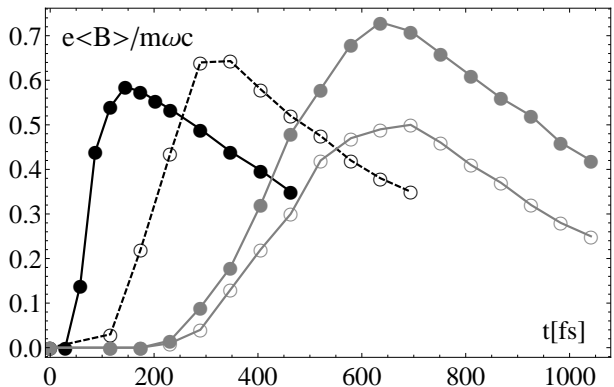}
\caption{The time dependence of the average magnetic field $B_z$ (0.1
approximately corresponds to 1\,kT) generated by a laser pulse with the
energy 0.5\,J and durations 30\,fs (in black), 120\,fs (open circles, dashed
line), and 300\,fs (in gray): the gray line with open circles shows the
magnetic field component in the $Z$ direction in the vicinity of the target
boundary in the third case.}
\label{fig4}
\end{figure}

We also studied how the magnetic field generation depends on the laser pulse
duration. We increased the laser pulse duration from 30\,fs to 120\,fs and
to 300\,fs while keeping laser energy constant (the laser intensity
respectively decreases four times and one order of magnitude). The magnetic
field evolution inside the target in the three cases is shown in
Fig.~\ref{fig4}.

It is clear that despite the laser intensity decrease, an increase of the
laser pulse duration results in an increase of the magnetic field amplitude.
In all the cases, the magnetic field amplitude reaches a maximum value after
the laser pulse terminates, i.e., the magnetic field growth times are indeed
equal to the laser pulse time duration. There is then a slower decrease of
the magnetic field average values with approximately the same characteristic
decay time of the order of 1.25\,ps. The maximum value of the magnetic field
is of the order of 7\,kT and can be achieved for the 300\,fs laser pulse
duration. We note that using long laser pulses (with duration $\gtrsim$
300\,fs) results in some nonuniformity of the generated magnetic field $Z$
component inside the target loop. It has a maximum near the foil center and
drops to the target boundary (cf.~two gray lines in Fig.~\ref{fig4} ).

\section{Discussion}

In this section, we present a simple qualitative analysis of the numerical
results, to increase their clarity and to relate them to a more realistic
situation of metal targets rather than the plasma model target used in the
numerical simulations. The magnetic field production and evolution can be
estimated by replacing the specifically shaped target (see Fig.~\ref{fig1})
with a series circuit of a capacitor and inductor (cf.~\cite{Santos-pop2018}).
Such a representation is valid for a quasistatic field if its characteristic
wavelength is much greater than the target size \cite{LL8-engl}. For the
driven target, two limit situations can be distinguished depending on the
ratio $\varkappa=\tau_L/\tau_*=\tau_Lc/P_*$ of the laser pulse duration
$\tau_L$ to the time of the electromagnetic pulse propagation $\tau_*$ along
the target perimeter $P_*$ at the speed of light $c$. For $\varkappa\ll1$,
the discharge once excited propagates as a wave in the target. The process
is strongly nonstationary and should be considered based on the full set of
Maxwell equations with the displacement current and corresponding dielectric
permittivity. In the opposite situation, $\varkappa\gg1$, the discharge
driven by the laser provides a quasistationary electron current source that
allows a simpler estimate for the magnetic field. The PIC simulations are
applicable in both regimes.

For the used target geometry, the contour length is $r\sim15\,\mu$m, which
corresponds to the time $\tau_*\sim50$\,fs, and the 30\,fs FWHM laser pulse
hence corresponds to the parameter $\varkappa\lesssim1$. An increase of the
pulse duration to 120\,fs yields $\varkappa\gg1$. Below, we use the
quasistatic approach to estimate the characteristic values, but we note that
for a more accurate description, the electromagnetic problem must be
considered. The oscillatory structure in Figs.~\ref{fig2} and \ref{fig3}
demonstrates the role of the wave processes for the parameters of interest.

The magnetic field generation process under irradiation of the target by an
intense relativistic laser pulse is governed predominantly by the target
properties, such as its geometric shape and the dielectric permittivity.
First, we consider the gap between the emitter and collector plates at the
bottom of the target in Fig.~\ref{fig1}. When the laser pulse irradiates the
emitter, electrons are accelerated inside the gap forward to the collector,
creating a charge density and a potential difference (electrostatic sheath)
inside the inter-plate gap. The formed sheath potential keeps most of the
ejected electrons at a distance of the order of the Debye length, which is
less than the gap thickness. For a laser field amplitude $a_0\sim8.5$, the
ponderomotive electron energy scaling \cite{Wilks-prl92} gives an effective
electron temperature $T_h\sim\sqrt(1+a_0^2)-1\sim7.6$\,MeV. Assuming that
$\sim30\%$ of the laser energy is deposited in the hot electrons (see, e.g.,
\cite{Roth-cern2017}), we can estimate their number as
$N_h\sim0.3E_{las}/(3T_h)$, i.e., $\sim10^{11}$ electrons. Distributing all
these electrons over the volume given by the Debye sheath thickness and a
hot spot with a radius of the order of $5\,\mu$m, we can estimate the
ejected electron density as $\sim10^{22}$\,cm$^{-3}$, which results in a
Debye length of $\sim0.1\,\mu$m. The positive potential of the emitter
created during laser irradiation of the target drives a replacement
(discharge) electrical current. This replacement current structure
propagates from the irradiated emitter along the target surface as a
high-frequency discharge wave and on the time scale of $\tau_*$ reaches the
collector to create a strong positive potential there. At this instant, it
disrupts the locking sheath potential, an acceleration potential appears in
the gap, the laser-heated electrons start to accelerate toward the
collector, and short circuit occurs. After that, the electrical current
enters the quasistatic regime. If the laser pulse lasts considerably longer
than $\tau_*$, almost all laser-generated hot electrons are not locked by
the sheath potential and participate directly in the quasistatic current.

The magnetic field strength in a quasistationary situation can be estimated
as $B\sim\mu_0I/2r$, where $r$ is the characteristic scale of the target
radius. Using the above estimate of the number of hot electrons, we can
calculate the maximum possible current $I\sim500$\,kA. This electrical
current is able to produce a magnetic field at the level of $B\sim20$\,kT.
But in the case of a short pulse, the quasistationary condition is not
completely satisfied, and the second plate cannot accept all the current
instantaneously. The efficiency of the magnetic field generation therefore
seems low. For a longer pulse, the quasistatic situation allows electrons to
form a steady circuit, and the resulting magnetic field seems closer to the
above estimate. The numerical difference in these estimates can appear from
estimation uncertainties in the target geometry, electron energies,
absorption efficiency, and other factors.

The magnetic field strength, its growth rate duration, and the efficiency of
generation demonstrated above reasonably correspond to the PIC simulations
with the collisionless plasma model target. After excitation, the magnetic
field relaxes. Because the simulations are collisionless, there is no ohmic
resistance in a plasma target, and magnetic field decay with the time scale
of about 1\,ps shown in Fig.~\ref{fig4} can differ from that in a real
experiment with a metal foil with a finite conductivity. The magnetic field
decay (Fig.~\ref{fig4}) is likely to be due to electromagnetic wave emission
from the target loop by a nonstationary electrical current. We note that
sharp kinks on the curved foil also contribute to such emissions as enhanced
localized sources of high-frequency energy leakage. The amplitudes of the
emission fields are small compared with the generated magnetic field
strength. Therefore, the characteristic decay time of the magnetic field is
longer than its growth rate duration. Some indication of the emission
mechanism is provided by the oscillatory structure in Fig.~\ref{fig3}.

In the case of a metal foil coil target, the magnetic field relaxation can
be qualitatively described by the solution of the equivalent LR circuit
equation
\begin{equation*}
V=RI+L\frac{dI}{dt},
\end{equation*}
where $V$ is the source voltage, $R$ is the target resistance, $I$ is the
electrical current in the target, and $L$ is its inductance. The relaxation
time is defined by $\tau_{LR}\sim L/R$ and $I\propto B\propto\exp(-t/\tau_{LR})$.
The target inductance estimated in SI units is $L\sim2\mu_0r\ln({2\pi r}/{a})
\sim6$\,pH. For the resistance calculation, a correct target cross section
where electrical current can flow should be taken into account. This cross
section is defined by the target skin depth in the high-frequency domain.
Correspondingly, it increases for the longer relaxation time scale. The
resistivity in copper is about $\rho\sim10^{17}$\,s$^{-1}$ in CGS units or
$\rho\sim2\times10^{-8}\,\mathrm\Omega$m and usually increases with
temperature \cite{Zaghloul-pop08}. On a time scale of 30\,fs, the surface
target layer corresponding to where a current propagates would be of the
order of or less than one tenth of a micron. Then $R\sim2\pi r\rho/S\sim
2\times10^{-3}\,\mathrm\Omega$, where $S\sim0.1\,\mu$m${}\times5\,\mu$m
and $\tau_{LR}\sim6$\,ps. We can therefore expect the magnetic field decay
to be several times slower than what is seen in the PIC simulations. For
more realistic calculations, the frequency and temperature dependence of the
resistivity should be carefully analyzed. We note that these estimates
applicable in the quasistatic limit are similar to those for capacitor-coil
targets \cite{Tikhonchuk-pre2017}, although the target geometry and all
characteristic times and lengths are quite different.

\section{Conclusion}

We have presented a new, rather simple scheme for generating ultra-strong
magnetic fields in specially designed targets under the action of ultrashort
intense table-top laser pulses. The performed 3D PIC simulations predict the
generation of a quasistationary magnetic field with a maximum amplitude up
to 7\,kT in a volume of about 100\,$\mu$m$^{3}$ generated by a 500\,mJ
femtosecond laser pulse, which is quite a strong field for such a moderate
laser energy. We studied the time evolution of the generated magnetic
fields. The qualitative estimates show that the time duration of the
magnetic field generated using a metal foil with finite conductivity would
be at least at the ten picosecond time scale, much longer than the driving
laser pulse.

\subsection*{Acknowledgments}
This work was supported by the Russian Foundation for Basic Research (Grant
No.~18-02-00452-a). Numerical simulations were
partially performed at the Joint Supercomputer Center of the Russian Academy
of Sciences and with resources of NRNU MEPhI High-Performance Computing
Center.


\end{document}